# Fine-grained Classification of Port Wine Stains Using Optical Coherence Tomography Angiography

Xiaofeng Deng, Defu Chen, Bowen Liu, Xiwan Zhang, Haixia Qiu, Wu Yuan, *Senior Member, IEEE*, Hongliang Ren, *Senior Member, IEEE*

*Abstract*—Accurate classification of port wine stains (PWS, vascular malformations present at birth), is critical for subsequent treatment planning. However, the current method of classifying PWS based on the external skin appearance rarely reflects the underlying angiopathological heterogeneity of PWS lesions, resulting in inconsistent outcomes with the common vascular-targeted photodynamic therapy (V-PDT) treatments. Conversely, optical coherence tomography angiography (OCTA) is an ideal tool for visualizing the vascular malformations of PWS. Previous studies have shown no significant correlation between OCTA quantitative metrics and the PWS subtypes determined by the current classification approach. This study proposes a new classification approach for PWS using both OCT and OCTA. By examining the hypodermic histopathology and vascular structure of PWS, we have devised a fine-grained classification method that subdivides PWS into five distinct types. To assess the angiopathological differences of various PWS subtypes, we have analyzed six metrics related to vascular morphology and depth information of PWS lesions. The five PWS types present significant differences across all metrics compared to the conventional subtypes. Our findings suggest that an angiopathology-based classification accurately reflects the heterogeneity in PWS lesions. This research marks the first attempt to classify PWS based on angiopathology, potentially guiding more effective subtyping and treatment strategies for PWS.

## I. INTRODUCTION

PORT wine stains (PWS) are congenital vascular malformations characterized by ectatic capillary vessels that manifest as reddish-purple discolorations on the skin. These lesions affect approximately 0.3% to 0.5% of newborns and persist throughout life [1]. PWS can lead to significant psychological distress, social stigmatization, and various physical complications, including hypertrophy, nodularity, and bleeding, particularly if left untreated [2]. Additionally, PWS can be associated with more severe complications, such as glaucoma [3], Sturge-Weber syndrome [4], which includes neurological issues like seizures and developmental delays, and Klippel-Trenaunay syndrome, which involves limb overgrowth and varicose veins [5]. Accurate diagnosis classifications and effective treatment are essential for cosmetic purposes, enhancing quality of life, and preventing secondary complications [6].

Vascular-targeted photodynamic therapy (V-PDT) has emerged as a prominent treatment modality for PWS [7-9]. V-PDT involves the administration of photosensitizer, which preferentially accumulates in the abnormal vasculature, followed by irradiation with a specific wavelength of light. It induces localized photochemical reactions that lead to vascular occlusion and subsequent lesion blanching. However, the clinical use of V-PDT in treating PWS has shown varying levels of effectiveness. While certain patients experience substantial lesion reduction, others see minimal to no improvements [10, 11]. In cases where V-PDT is ineffective, the phototoxicity associated with residual photosensitizer and light exposure can exacerbate the condition, resulting in thicker macules or nodules that necessitate surgical intervention and prolonged postoperative monitoring [12]. The variability in V-PDT efficacy is hypothesized to be associated with angiopathological differences in the vascular lesions, including vessel diameter, depth, and the extent of vertical growth. For instance, studies have indicated that PWS lesions with larger and deeper vessels may exhibit a poorer response to V-PDT due to insufficient penetration of the photosensitizer and light [13, 14]. Conversely, lesions with a higher density of smaller vessels may respond more favorably due to more effective accumulation and activation of the photosensitizer [15, 16].

Currently, clinical diagnosis and assessment of PWS are primarily based on visual inspection of the skin's surface, considering factors like color, lesion size, and the presence of skin thickening or nodularity [17]. Dermatologists typically classify PWS lesions into three types—pink, purplish red (PR), and thickened—based on their color and morphological changes, such as hypertrophy and nodularity, to formulate treatment strategies [18]. However, these subtypes do not

This work was supported by the Hong Kong Research Grants Council (CRF C4026-21G, GRF 14216022, ECS24211020, GRF14203821, GRF14216222), National Natural Science Foundation of China (62227823, 62205025), the Science, Technology and Innovation Commission (STIC) of Shenzhen Municipality (SGDX20210823103535014, SGDX20220530111005039). (Xiaofeng Deng, Defu Chen and Bowen Liu are co-first authors) (Haixia Qiu, Wu Yuan and Hongliang Ren are co-corresponding authors).

Xiaofeng Deng and Hongliang Ren are with the Department of Electronic Engineering, The Chinese University of Hong Kong, Hong Kong, China (e-mail: dengxf@link.cuhk.edu.hk; hlren@ieee.org).

Defu Chen and Xiwan Zhang are with the School of Medical Technology, Beijing Institute of Technology, Beijing 100081, China (e-mail: defu@bit.edu.cn; 3220235085@bit.edu.cn).

Bowen Liu and Wu Yuan are with the Department of Biomedical Engineering, The Chinese University of Hong Kong, Hong Kong, China (e-mail: luke4west@outlook.com; wyuan@cuhk.edu.hk).

Haixia Qiu is with the Department of Laser Medicine, the First Medical Centre, Chinese PLA General Hospital, Beijing 100853, China (e-mail: qiuhxref@126.com).

adequately reflect the underlying histopathology and vascular structure, which are critical for phenotyping and optimizing V-PDT efficacy. For example, the color of PWS lesions is influenced not only by the degree of expansion of malformed blood vessels, but also by factors such as the melanin content in the epidermis, epidermal thickness, and the depth of blood vessels [19]. In other words, the current PWS classification scheme, which relies on the aforementioned appearance features of the skin, fails to capture the complex angiopathology of PWS lesions, leading to the unsatisfactory treatment effect in V-PDT. Accurate diagnosis and effective treatment of PWS require visualizing these abnormal blood vessels within the affected tissues. While skin biopsy remains the gold standard for histopathological evaluation, the invasive nature in histological analysis of PWS makes it impractical for routine clinical use [20]. Therefore, there is an urgent need for non-invasive optical imaging technologies that can provide enhanced insights into depicting detailed angiopathology of PWS lesions to guide treatment decisions and improve clinical outcomes.

Optical Coherence Tomography (OCT) is an emerging non-invasive technique that offers high-resolution 3D imaging of biological tissues [21-23]. As its functional extension, OCT angiography (OCTA) enables the visualization of blood vessels by detecting motion contrast from flowing red blood cells, offering a detailed 3D view of the vascular network without the need for contrast agents [24, 25]. With the advantages of non-contact, label-free, high-resolution, rapid and 3D imaging capability, OCT and OCTA have been instrumental and widely adopted in clinical applications [26-30]. Recently, OCT and OCTA have also shown great application potential in dermatology for various diagnostic purposes, including assessing inflammatory skin diseases like neurodermatitis and psoriasis [31, 32], and monitoring skin cancers such as cutaneous melanoma [33, 34]. Recent studies have demonstrated the utility and reliability of OCTA in imaging and quantifying PWS [35-38]. However, existing 2D and 3D quantitative OCTA metrics [37, 38] only focused on differentiating PWS lesions from healthy normal skin tissue, with a limited exploration into the variability within PWS lesions of different clinical appearances. Given this, quantitative OCTA metrics were applied to patients with PR and thickened types of PWS diagnosed by clinical skin appearance but found no significant differences in vascular diameter, vascular density and vessel depth [36]. Despite the advancements in quantitative OCTA analysis, the lack of significant quantitative differences across PWS subtypes defined by skin appearance suggests that visual classifications do not adequately capture the underlying angiopathological heterogeneity of PWS lesions. It remains unclear whether PWS can be classified according to the angiopathological differences of the lesions instead of the features of the skin appearance. This hypothesis emphasizes the need for a more nuanced understanding of PWS angiopathology to inform treatment strategies.

This study aims to develop a novel method of classifying PWS based on the angiopathology of lesions. To achieve this, we employ OCT and OCTA as optical biopsy techniques to visualize the hypodermic histopathology and vascular structure of PWS lesions. Subsequently, we establish a framework for quantifying angiopathological characteristics by evaluating six metrics that capture vascular morphology and lesion depth in OCT and OCTA images. We then develop a fine-grained classification method that divides PWS data into five types using labeled pairs of PWS data and corresponding contralateral normal skin data. Our results demonstrate that compared to the conventional classification based solely on visual observation of facial skin appearance, the proposed fine-grained classification method exhibits more significant differences between various types across all six quantitative metrics. To the best of our knowledge, this is the first classification approach that unveils the diversity of angiopathology from hypodermic lesions, thereby investigating the heterogeneity of PWS. Our findings can potentially guide the clinical diagnosis and treatment of PWS.

## II. METHODS

### A. Subject Population and Data Collection

Fifty-seven individuals with PWS were enrolled in the Department of Laser Medicine at the Chinese PLA General Hospital's First Medical Center. The study was approved by the hospital's Ethics Committee, and written informed consent was obtained from each participant or their legal guardian for those under 18. Notably, all participants were free from concurrent conditions such as Sturge-Weber syndrome or glaucoma.

OCT imaging was performed on both the PWS lesion and unaffected contralateral skin on the zygomatic region of each patient. Depending on the distribution of PWS lesions, 4 ~ 6 PWS regions of interest (ROI) and their corresponding contralateral regions were scanned for each patient. Each region was scanned twice for repeatability. A handheld SS-OCT device (Tianjin HoriMed Medical Technology, China) was used to capture the OCT data, featuring a 200 kHz A-line scan rate and a central wavelength of 1310 nm, which enabled an axial resolution of approximately 15 μm. The probe beam was focused on the skin surface, resulting in a lateral resolution of around 10 μm at the superficial dermis level. The OCT data consisted of 1500 repeated B-scans at 500 tomographic positions, with each B-scan comprising 500 × 1024 voxels in the z-x plane. This corresponded to a total acquisition time of 3.75 s. The OCT volume covered a 3D scanning range of 4 × 4 × 2.1 mm.

A custom holder with a glass window was firmly attached to the handheld OCT probe to ensure stability during imaging (see Fig. 1b). The holder was designed to contact the skin, reducing motion artifacts. Additionally, a layer of glycerin was applied to the glass window to optimize the refractive index, minimizing surface reflections and enhancing light penetration into the skin.

### B. OCTA Processing and Quantification Framework

To mitigate the influence of motion artifacts resulting from the micro movement of subjects' faces, we employed an inter-frame registration method based on correlation features between three repeated B-frames of the same location [39]. Subsequently, a cross-sectional OCT image was generated by averaging the three repeated B-frames, followed by applying a logarithmic transformation and normalizing the intensity values of each pixel to a range of 0 to 1. Additionally, a graph search method based on Dijkstra's shortest path algorithm was applied to find the upper surface of skin in each cross-sectional OCT image [40].

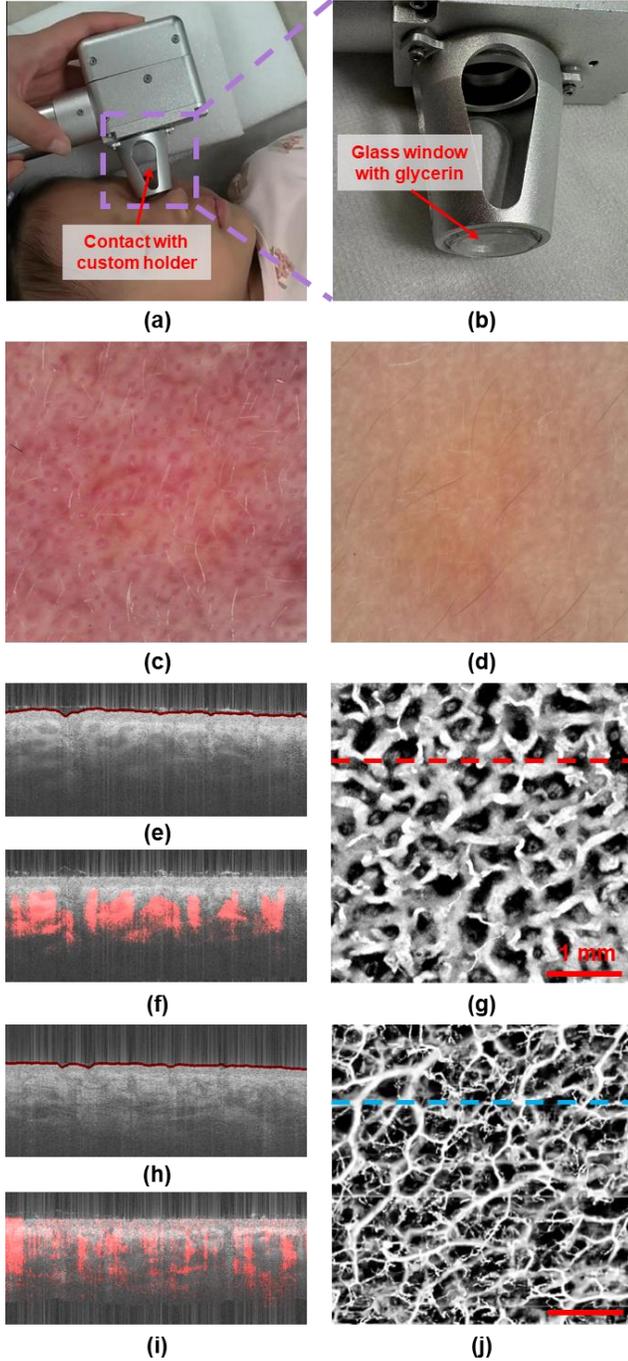

Fig. 1. Representative clinical data collection scene, dermoscopy images, OCT and OCTA images for PWS lesion and contralateral normal skin of a patient. (a) Photo shows a clinician performing OCT/OCTA detection on a PWS patient using handheld OCT probe. (b) Enlarged close-up image of the custom holder. (c) Dermoscopy image of PWS lesion area. (d) Dermoscopy image of contralateral normal skin. (e) Cross-sectional OCT image with an identified upper surface of PWS lesion area. (f) Cross-sectional OCT/OCTA overlapped the image of the PWS lesion area and flattened it according to the identified upper surface. (g) En-face OCTA image of PWS lesion area. Red dashed line indicates the cross-sectional location of (e) and (f). (h) Cross-sectional OCT image with identified upper surface of contralateral normal skin area. (i) Cross-sectional OCT/OCTA overlapped image of contralateral normal skin area, flattened according to the identified upper surface. (j) En-face OCTA image of contralateral normal skin area. Blue dashed line indicates the cross-sectional location of (h) and (i).

The ID-OCTA algorithm [41, 42] generated cross-sectional OCTA images by analyzing temporal changes between the three repeated OCT B-frames. Then, cross-sectional OCT and OCTA images of the same location were flattened along the determined upper surface and overlaid to facilitate subsequent calculations of blood flow depth. Figure 1 shows a clinical OCT data collection scene and a multi-modality case exhibition of a patient with PR-type PWS, including dermoscopy images (magnification: 50X) and OCT/OCTA images.

To obtain a comprehensive representation of the vascular network, we performed maximum value projection (MVP) on the 3D OCTA volume constructed by stacking all the cross-sectional OCTA images [43]. This process yielded an en-face OCTA image and the depth information of each maximum value located in the 3D coordinate, which was further utilized for quantitative analysis.

To quantitatively evaluate the angiopathological differences of PWS lesions, a customized quantification framework was employed to acquire six desired metrics containing vascular morphology and depth information. Details of this framework are shown in Figure 2.

1). Considering that PWS lesions have more wrinkled skin surface than normal skin, surface tortuosity (ST) was introduced to characterize the degree of irregularity and protuberance of the skin surface. ST was computed by evaluating the standard deviation of the depth values corresponding to the identified skin surface in cross-sectional OCT images (Fig. 2a).

2). To enhance the accuracy of subsequent analyses, the raw en-face OCTA images (Fig. 2b) underwent denoising using the BM3D method [44] and binary operation using the Phansalkar threshold [45]. The resulting clean binary vascular maps (Fig. 2c) served as the basis for calculating the indexes of vessel density (VD), vessel caliber (VC), and vessel complexity (VX), which are widely employed in ophthalmology and dermatology [36, 46, 47].

3). The binary vascular maps were color-coded according to the depth value of each pixel below the identified upper surface, resulting in the generation of vascular-depth maps (Fig. 2d). To quantify the vessel depth difference between PWS lesions and normal skin, the mean depth of vessels (MDV) was determined by calculating the average depth value of all non-zero pixels within a vascular-depth map.

4). The vascular-depth map was skeletonized, and the two endpoints of each vessel were marked (Fig. 2e). The index of the degree of vertical growth (DVG) was calculated in the following equation:

$$DVG = \frac{\Sigma \frac{\nabla(m_i, n_i)}{L(m_i, n_i)}}{I} \quad (1)$$

Here, $m_i$ and $n_i$ represented 2 endpoints of the $i^{th}$ vessel, while $I$ denoted the total number of vessels in the skeletonized vascular-depth map. $\nabla(m_i, n_i)$ and $L(m_i, n_i)$ indicated the depth drop along with the vessel and vessel length of the $i^{th}$ vessel, respectively. DVG was proposed to reflect the tendency of columnar inclination of blood vessels within a vascular-depth map.

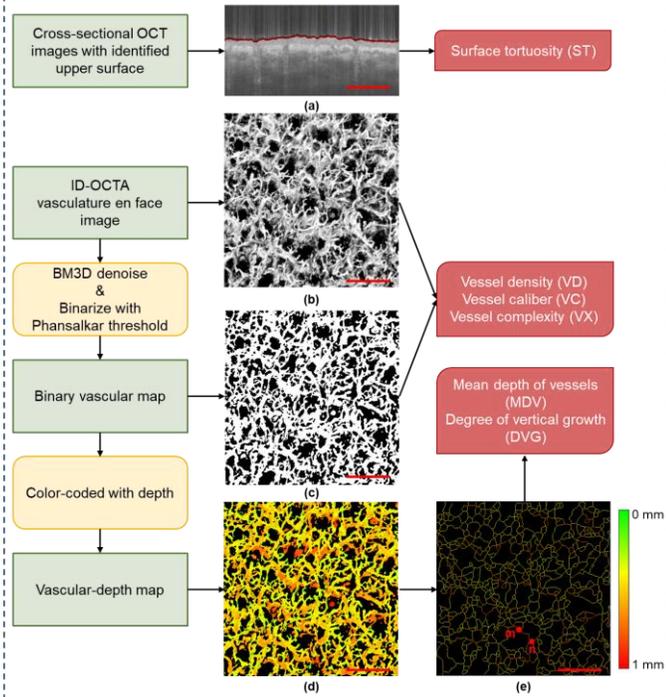

Fig. 2. Illustration of the proposed quantification framework for 6 metrics of vascular morphology and depth information. (a) Cross-sectional OCT image with identified upper surface. (b) En-face OCTA image. (c) Binary vascular map. (d) Vascular-depth map. (e) Skeletonized vascular-depth map. (d) and (e) are color-coded with the depth value of each pixel from the skin surface.

### C. Workflow of Fine-grained Classification of PWS

In this study, we address the challenge by learning more refined embeddings of vasculature beyond the coarse, binary labels (C: control group of normal skin data, P: PWS lesions) provided by clinical data. Relying solely on binary labels introduces latent space divisions dominated by hyperplanes, which are not conducive to the nuanced mining of vasculature information. Subsequently, we apply clustering techniques to the learned latent space. To facilitate interpretation, we designate the cluster of normal samples as a reference and rank the remaining clusters based on their Euclidean distance from this reference cluster. This process results in an annotated latent space. Utilizing this annotated latent space, we develop a k-nearest neighbors (k-NN) model for data retrieval. This method maps the vasculature information of PWS lesions to a low-dimensional dense embedding, providing a fine-grained description. Figure 3 delineates the workflow of our methodology. The following sections will elaborate on each component of our approach in detail.

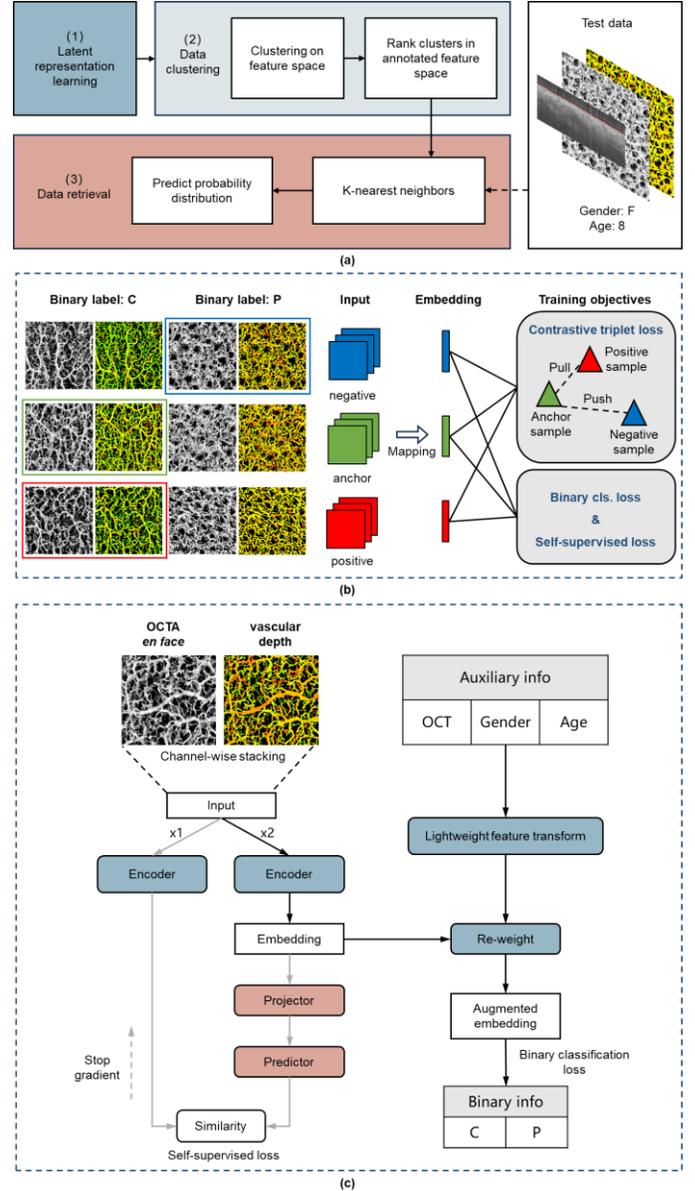

Fig. 3. Workflow of fine-grained classification for PWS. (a) The pipeline of the proposed fined-grained classification of PWS includes a latent representation learning process for feature extraction, data clustering analysis for annotating the learned latent space, and data retrieval under a k-NN model. (b) Illustration of the process of latent representation learning. (c) Implementation of the self-supervised loss and the binary classification loss.

#### 1) Learning procedure

The first part of the pipeline of fined-grained classification (Fig. 3a) is the latent representation learning process, which is used to select the most informative features from OCTA images and other information. The extractor can learn finer features under the supervision of coarse-grained clinical labels. Then, data clustering is conducted on the latent space, and each cluster is ranked, and a new label is assigned based on the Euclidean distance relationship. The annotated feature space is utilized as the holdout set during the data retrieval. The probabilities of input test data clustered into each category are then predicted using nearest-neighbor voting.

In the procedure of latent representation learning (Fig. 3b),

we define the channel-wise concatenation of the enface OCTA image and its vascular-depth map as input. The input is mapped into the corresponding embedding by the encoder, which is then passed into different loss functions for optimization. There are three loss functions in total. The calculation of contrastive triplet loss [48, 49] requires the construction of triplets at the input end, while the other two losses are calculated for each element in the triplet.

### 2) Contrastive triplet loss

We aim to achieve mutual supervision for contrastive triplet loss by leveraging the correlations among multiple images corresponding to the same patient. In detail, OCTA data from the same patient are divided into two classes with clinical binary labels: C and P. The triplet is constructed as follows: one pair of en-face images and a vascular-depth image of a patient are randomly selected and denoted as the Anchor sample (denoted as $x_a$). Subsequently, another pair of data from the same class with an Anchor sample is selected and denoted as a Positive sample (denoted as $x_p$), and a pair of data from the other class is selected and denoted as Negative (denoted as $x_n$).

For the three elements (Anchor, Positive, Negative) in the triplet, after feature extraction by the feature encoder, we obtain embeddings denoted as $e_a$, $e_p$, $e_n$, respectively. As shown in Fig.3b, a contrastive triplet loss is leveraged to minimize the distance between $e_a$ and $e_p$, and maximize the distance between $e_a$ and $e_n$, ensuring a minimum margin $\alpha$ between the distance from $e_a$ to $e_n$ and $e_a$ to $e_p$. The calculation formula is as follows:

$$L_{triplet} = \sum max \left( \left\| e_a^i - e_p^i \right\|_2^2 - \left\| e_a^i - e_n^i \right\|_2^2 + \alpha, 0 \right) \quad (2)$$

where $i$ represents the number of triplets. The distance is measured by Euclidean distance. Finally, by minimizing the distance between Positive and Anchor samples while maximizing the distance between Anchor samples and Negative samples, the contrastive triplet loss learns more discriminative concepts of vasculature and extracts disease-related information shared among lesions from different skin locations of one patient.

### 3) Self-supervised loss

Inspired by Grafit [50], we can utilize the supervision offered by multiple data augmentations of the same input instance. Self-supervised learning enables us to tap into and utilize information beyond the coarse binary labels, leading to finer embeddings by emphasizing instance discrimination. Thus, based on SimSiam [51], we process multiple different augmented views of the input using the same encoder network plus the projector. We chose ResNet18 [46] and Multi-Layer Perceptron (MLP) in the experimental settings, respectively. We arbitrarily select one embedding as the query among the multiple embeddings obtained and perform pairwise analysis with the remaining embeddings. Specifically, a prediction MLP, $h$, is employed to process one set of embeddings, while the other set is subjected to a stop-gradient operation. The model's primary objective is to maximize the similarity between the two resulting embeddings, thereby fostering a strong correlation between them. In detail, denoting the input as $x$, the encoder plus the projector as $f$, and two output vectors as $p = h(f(x))$ and $z = f(x)$, the self-supervised loss is calculated as follows:

$$L_{sp} = \frac{1}{T-1} \sum_{1 \leq i \neq q \leq T} \left[ \frac{1}{2} D(p_q, z_i) + \frac{1}{2} D(p_i, z_q) \right] \quad (3)$$

Each term in the self-supervised instance loss function $L_{sp}$ is a symmetric loss, where the subscript $q$ corresponds to the query embedding, and $T$ denotes the number of augmented views for one input. The similarity function $D$ is negative cosine similarity, computed as follows.

$$D(p_*, z_*) = -\frac{p_*}{\|p_*\|_2} \cdot \frac{z_*}{\|z_*\|_2} \quad (4)$$

where $\|\sim\|_2$ is $l_2$-norm. Additionally, following the setup in SimSiam [51], the implementation of the similarity function $D$ under the experiment setting is described below.

$$D(p_*, z_*) = D(p_*, \text{stopgrad}(z_*)) \quad (5)$$

where stopgrad(~) denotes the stop-gradient operation when back-propagating to update the network parameters.

Drawing on the success of Siamese neural networks in unsupervised visual learning [52-54], our approach utilizes a straightforward Siamese architecture. This design fosters the learning of finer representations based on the "free" annotations from instance-level discrimination without needing other sample pairs or large batch sizes. Leveraging these benefits, we incorporate this embedding consistency with the binary classification loss, emphasizing instance discrimination under coarse-grained supervision.

### 4) Binary classification loss

After obtaining the embedding from the encoder based on the OCTA information, other information used for supervised learning includes OCT image (with identified skin surface), gender, and age. Firstly, for input information processing, we randomly zeroed some of the frames of the OCT videos with probability $p$ ($p = 0.1$ in experiments), retaining only a small number of frames. The zeroed frames are chosen independently for each forward call and are sampled from a Bernoulli distribution. Gender and age correspond to binary variables (0 for Female, 1 for Male) and continuous variables (in years). Next, we employ a lightweight network (MobileNetV2 [55] in experiments) to process the OCT input. Two fully connected layers handle the gender and age inputs separately. Finally, leveraging the output embeddings as complementary information, we utilize cross-attention to weight the OCTA embedding, obtaining the augmented embedding. After one fully connected layer, the output logit is used for binary classification loss. The calculation formula is as follows.

$$L_{bc} = \frac{\sum -[y_i^* \cdot \log \sigma(y_i) + (1-y_i^*) \cdot \log(1-\sigma(y_i))]}{N} \quad (6)$$

where $y_i$ and $y_i^*$ denote the output logit and the binary label of the $i$ th input, and $\sigma(\sim)$ denotes the sigmoid function.

*5) Constructing predictive k-NN model for data retrieval*

Upon completing the encoder's training, we execute K-means++ clustering [56] on the feature space and obtain five clusters of PWS samples in addition to one cluster of normal samples. The feature space for clustering only contains samples corresponding to the lesion skin, as we do not need to subdivide the samples corresponding to normal skin. For further interpretation, we take the cluster of normal samples as a reference, and rank the rest of the clusters according to the Euclidean distance between their cluster centers and the reference cluster center, resulting in an annotated feature space. After annotating the feature space using the clustering results, the features and their annotations will serve as the hold-set for constructing the predictive k-NN model. During the data retrieval, it is only necessary to perform feature transformation on the query data and compare its similarity with each cluster, ultimately determining which cluster it belongs to. In this way, we map the vasculature information to a low-dimensional dense embedding, serving as a fined-grained classification scale for PWS.

### D. Statistical Analyses

Statistical analyses for quantitative results were performed with SPSS, version 23.0 (SPSS Inc., Chicago, IL, USA), using one-way analysis of variance (ANOVA) with post-hoc Tukey honest significant difference tests for multiple comparisons. Statistical significance was defined by p-value < 0.05.

## III. RESULTS

### A. PWS Subtypes Comparison Between Appearance-based and Proposed Fine-grained Classification

We acquired 525 OCT volumes from PWS lesions and 525 OCT volumes from corresponding contralateral normal skin areas. Dermatology experts divided 525 PWS data into pink type (marked as PI), purplish red type (marked as PR) and thickened type (marked as TH) with 190, 285 and 50 cases, respectively, according to the clinical classification based on the skin appearance characteristics [17, 18]. Fig. 4 shows typical facial photos and corresponding OCTA images of three types of PWS patients diagnosed with skin appearance.

The proposed fine-grained classification network accurately distinguished normal (marked as type 0) and PWS data and classified PWS data into 5 categories (marked as type 1-5). Fig. 5a&b demonstrate the typical OCTA images of each type classified by the proposed network. Additionally, we used uniform manifold approximation and projection (UMAP) to visualize the difference in data distribution between the two classification results (Fig. 5c-d). Here, various types classified by the proposed network are sorted according to the average distance from the normal data in UMAP. In other words, type 1 is closer to normal data, and type 5 is further from normal data than other types. The dice similarity coefficient ($DSC = 2\frac{A \cap B}{A+B}$) is leveraged to quantitatively calculate the similarity of two groups and data distribution overlap of various PWS types. The result (Fig. 5e) shows that two classification methods consistently identify normal and PWS data. However, all types of PWS data classified by skin appearance exist in almost every disease type classified by the proposed network. These findings indicate significant differences between the skin appearance-based method and the proposed fine-grained classification in classifying PWS data.

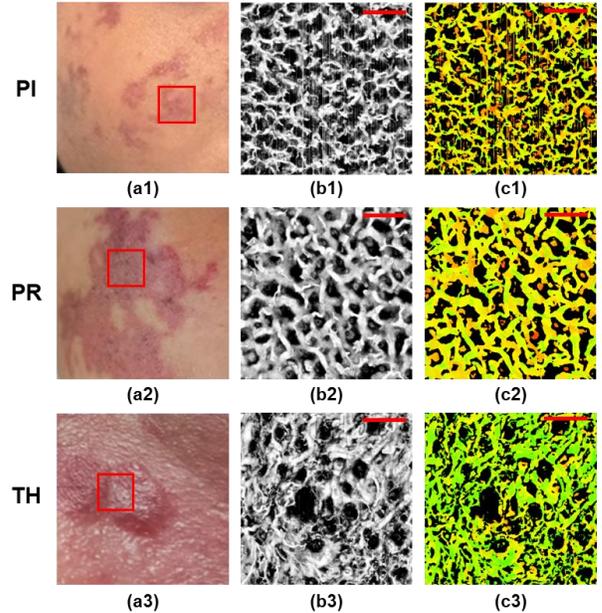

Fig. 4. Representative facial photos and OCTA images of PWS patients diagnosed by skin appearance. (a1-c1) Representative facial photo, en-face OCTA image and vascular-depth image of PI patient. (a2-c2) Representative facial photo, en-face OCTA image and vascular-depth image of PR type patient. (a3-c3) Representative facial photo, en-face OCTA image and vascular-depth image of TH patient.

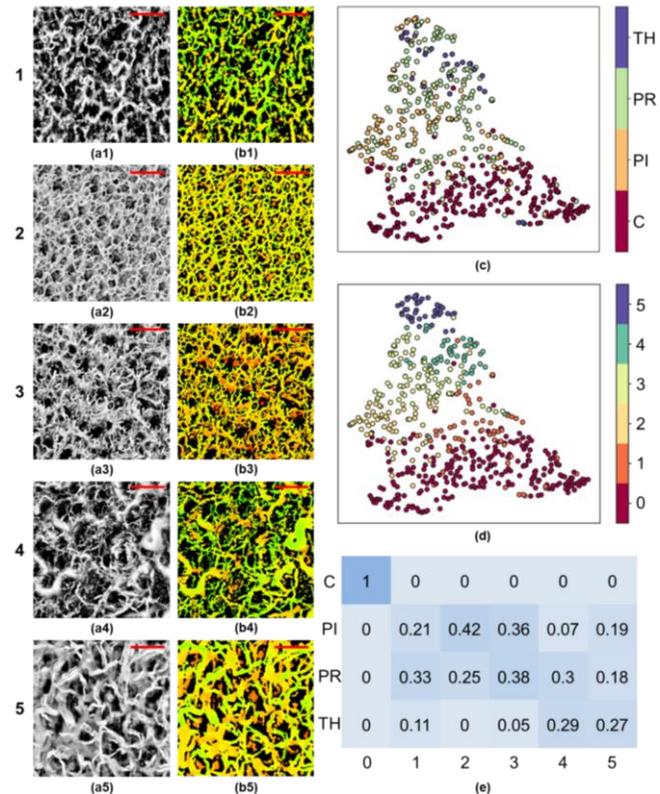

Fig. 5. Representative OCTA images of PWS types classified by proposed fine-grained network and data distribution comparison of two classification methods. (a1-b1) Representative en-face OCTA image and vascular-depth image of type 1 patient. (a2-b2) Representative en-

face OCTA image and vascular-depth image of type 3 patient. (a3-b3) Representative en-face OCTA image and vascular-depth image of type 1 patient. (a4-b4) Representative en-face OCTA image and vascular-depth image of type 4 patient. (a5-b5) Representative en-face OCTA image and vascular-depth image of type 5 patient. (c) Data distribution UMAP of PWS types classified by skin appearance-based method. (d) Data distribution UMAP of PWS types classified by the proposed fine-grained network. (e), Dice similarity coefficient matrix between two classification methods.

## B. Quantification of Vascular Morphology and Depth Information for Different Classification Results

By performing the previously described angiopathological quantification framework, we quantified all three types of PWS subtyped by skin appearance-based classification (SAC, including PI, PR, and TH) and five types of PWS subtyped by proposed fine-grained classification (FGC, including type 1-5) in 6 metrics related to vascular morphology and depth information of the lesions.

In the quantification process, we first calculated the exact value of each metric between one PWS data set and its corresponding contralateral normal data set. Then, the percentage change value of PWS data relative to the normal data was calculated for statistics. This operation ensured the reliability of quantification by removing the influence of other variables, such as skin position and individual differences between patients.

We obtained the mean value and standard deviation value of each metric for each subtype, then performed ANOVA and post-hoc analysis. Statistical difference analysis results are shown in Table I-IV.

TABLE I
ONE-WAY ANOVA ANALYSIS OF EVALUATION METRICS FOR PWS TYPES CLASSIFIED BY SKIN APPEARANCE

| Metrics | PI | PR | TH | p-value |
|---|---|---|---|---|
| No. of data | 190 | 285 | 50 | - |
| VD (%) | 3.97 ± 4.21 | 2.90 ± 3.44 | 3.06 ± 5.40 | 0.014 |
| VC (%) | 23.95 ± 13.72 | 28.78 ± 12.44 | 30.13 ± 15.27 | < 0.001 |
| VX (%) | -21.41 ± 15.12 | -24.60 ± 18.76 | -23.88 ± 16.65 | 0.143 |
| ST (%) | 8.60 ± 13.18 | 9.52 ± 15.56 | 69.41 ± 21.70 | <0.001 |
| MDV (%) | 3.09 ± 4.08 | 2.52 ± 3.02 | 3.22 ± 2.06 | 0.127 |
| DVG (%) | -6.23 ± 5.50 | -6.86 ± 6.02 | -5.36 ± 5.47 | 0.179 |

Values are expressed as mean ± standard deviation.

TABLE II
POST-HOC ANALYSIS OF THE DIFFERENT TYPES OF PWS CLASSIFIED BY SKIN APPEARANCE

| Metrics | PI vs PR | PI vs TH | PR vs. TH |
|---|---|---|---|
| VD | 0.011* | 0.316 | 0.962 |
| VC | < 0.001* | 0.009* | 0.783 |
| VX | 0.122 | 0.643 | 0.960 |
| ST | 0.801 | <0.001* | <0.001* |
| MDV | 0.169 | 0.968 | 0.367 |
| DVG | 0.476 | 0.611 | 0.210 |

* means p-value < 0.05.

TABLE III
ONE-WAY ANOVA ANALYSIS OF EVALUATION METRICS FOR PWS TYPES CLASSIFIED BY PROPOSED FINE-GRAINED CLASSIFICATION

| Metrics | Type 1 | Type 2 | Type 3 | Type 4 | Type 5 | p-value |
|---|---|---|---|---|---|---|
| No. of data | 102 | 113 | 150 | 85 | 75 | - |
| VD (%) | -6.19 ± 1.93 | 12.77 ± 5.97 | 4.87 ± 1.31 | -4.63 ± 1.90 | 7.80 ± 2.50 | <0.001 |
| VC (%) | 12.25 ± 3.08 | 11.35 ± 2.95 | 28.12 ± 4.49 | 36.04 ± 6.45 | 58.93 ± 8.50 | <0.001 |
| VX (%) | -6.47 ± 2.39 | -26.38 ± 8.35 | -23.46 ± 6.09 | -21.22 ± 4.03 | -43.79 ± 8.69 | <0.001 |
| ST (%) | 2.63 ± 1.26 | 6.18 ± 2.17 | 10.72 ± 2.65 | 29.06 ± 6.30 | 36.95 ± 8.14 | <0.001 |
| MDV (%) | 0.08 ± 0.11 | 2.57 ± 0.80 | 6.65 ± 1.52 | -2.05 ± 1.01 | 4.58 ± 1.27 | <0.001 |
| DVG (%) | -5.52 ± 1.30 | -15.49 ± 4.53 | -10.6 ± 2.83 | 6.65 ± 2.09 | -0.91 ± 0.30 | <0.001 |

Values are expressed as mean ± standard deviation.

TABLE IV
POST-HOC ANALYSIS OF THE DIFFERENT TYPES OF PWS CLASSIFIED BY PROPOSED FINE-GRAINED CLASSIFICATION

| Metrics | 1 vs 2 | 1 vs 3 | 1 vs 4 | 1 vs 5 | 2 vs 3 | 2 vs 4 | 2 vs 5 | 3 vs 4 | 3 vs 5 | 4 vs 5 |
|---|---|---|---|---|---|---|---|---|---|---|
| VD | <0.001* | <0.001* | 0.009* | <0.001* | <0.001* | <0.001* | <0.001* | <0.001* | <0.001* | <0.001* |
| VC | 0.704 | <0.001* | <0.001* | <0.001* | <0.001* | <0.001* | <0.001* | <0.001* | <0.001* | <0.001* |
| VX | <0.001* | <0.001* | <0.001* | <0.001* | 0.002* | <0.001* | <0.001* | 0.071 | <0.001* | <0.001* |
| ST | <0.001* | <0.001* | <0.001* | <0.001* | <0.001* | <0.001* | <0.001* | <0.001* | <0.001* | <0.001* |
| MDV | <0.001* | <0.001* | <0.001* | <0.001* | <0.001* | <0.001* | <0.001* | <0.001* | <0.001* | <0.001* |
| DVG | <0.001* | <0.001* | <0.001* | <0.001* | <0.001* | <0.001* | <0.001* | <0.001* | <0.001* | <0.001* |

* means p-value < 0.05.

Figure 6 presents the histogram of all subtypes from two classification methods in 6 metrics. For metrics characterizing vascular morphology of PWS lesions (VD, VC and VX), slight differences are observed between three types classified by SAC (PI vs. PR in VD, PI vs PR and PI vs TH in VC). However, 5 types of PWS data classified by FGC show apparent variety (Fig. 6a-c). Specifically, the VD of PI, PR and PR was approximately ~3% higher than the normal skin (Fig. 6a). Nevertheless, for the classification results of FGC (Fig. 6a), type 1 and type 4 PWS have lower VD compared to the normal skin (mean ± std: -6.19% ± 1.93% and -4.63% ± 1.90%, respectively), while type 2, type 3 and type 5 PWS have higher VD (12.77% ± 5.97%, 4.87% ± 1.31% and 7.80% ± 2.50%, respectively). PWS lesion contains ectatic blood vessels (Fig. 6b), while PR and TH cannot be distinguished by the difference of VC (28.78% ± 12.44% vs 30.13% ± 15.27%, p-value = 0.783). Although there is no significant difference between type 1 and type 2 PWS in VC, the dilation of vessels shows an increased trend in the five PWS types classified by FGC. Type 2 and type 5 PWS have minimal and maximal increases of 11.35 ± 2.95% and 58.93 ± 8.50% in VC, respectively. Moreover, the decrease of VX is observed in all PWS data (Fig. 6c), while significant differences are measured only among PWS types classified by FGC. In particular, type 1 and type 5 PWS show minimal and maximal decreases of 6.47% ± 2.39% and 43.79% ± 8.69% in VX, respectively. These results illustrate that it is hard to distinguish PI, PR, and TH through various OCTA metrics that quantify vascular morphology. However, the proposed fine-grained classification network can classify PWS data by fully unearthing vascular morphology features within OCTA images.

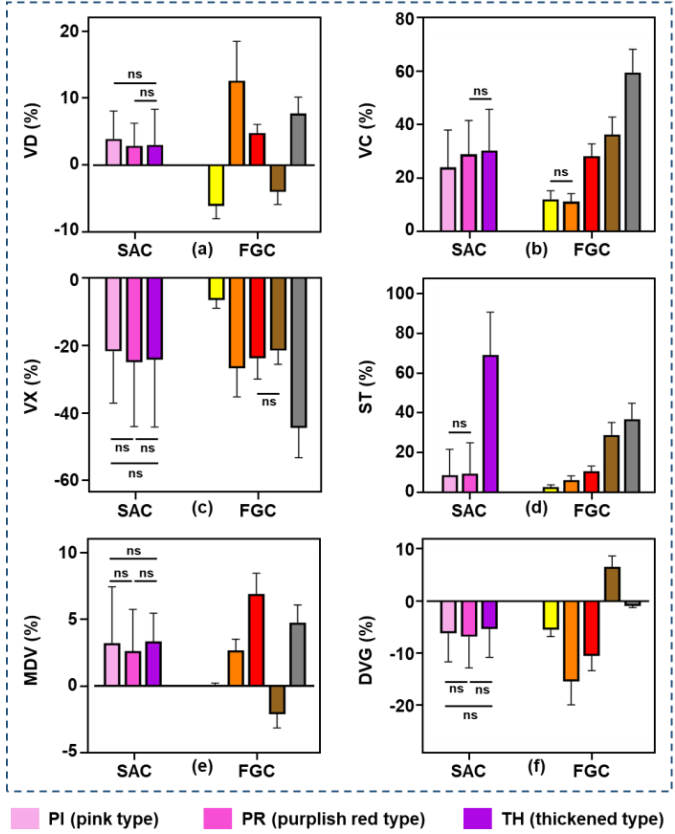

Fig. 6. Vascular and depth features quantification between groups of PWS types classified by two methods, using VD (a), VC (b), VX (c), ST (d), MDV (f) and DVG (g). SAC refers to skin appearance-based classification. FGC refers to fine-grained classification. ns means no statistical difference (p-value > 0.05).

We also assessed three metrics that contain depth information on the PWS lesions: ST, MDV and DVG (Fig. 6d-f). A significant increase is observed in the thickened type of PWS (69.41% ± 21.70%) compared to the other two types of PWS classified by SAC (Fig. 6d). This is a natural result because the swelling and roughness of the skin surface leads to the diagnosis of thickened type in SAC. An increase trend of ST is also found in five types of PWS classified by FGC. Additionally, the mean depth of blood vessels increases in all three types of PWS classified by SAC (3.09% ± 4.08%, 2.52% ± 3.02% and 3.22% ± 2.06%, respectively). Still, no significant difference is observed (Fig. 6e). In the types classified by FGC, the MDV of type 1 PWS remains unchanged compared to the normal data (0.08% ± 0.11%). When the other 3 types of PWS have an increase in MDV, only type 4 PWS shows a decline of 2.05% ± 1.01%. Type 3 PWS has a maximal increase of 6.65% ± 1.52% in MDV. Next, we measured the DVG of all PWS data (Fig. 6f). PWS types classified by SAC are generally lower than the normal control, while PWS types classified by FGC show apparent differences. Quantitatively, type 2 PWS has the lowest DVG compared to other types (-15.49% ± 4.53%). This is probably because type 2 PWS has a flatter vascular network (Fig. 5b2). Various from the overall decline in other types, the type 4 PWS shows a remarkable increase in DVG (6.65% ± 2.09%).

Notably, the standard deviation of all six metrics in types classified by SAC is higher than in types classified by FGC. This phenomenon demonstrates a greater intra-class variation of vascular morphology and depth features in the classification based on skin appearance (color and thickness). These results indicate that compared to the conventional skin appearance-based diagnosis method, the proposed fine-grained classification network can better reflect the intrinsic angiopathological heterogeneity of the PWS lesions by synthesizing the vascular morphology and depth information within OCTA and OCT images.

## IV. DISCUSSION

The abnormality of angiopathological features in the lesion area leads to deleterious vascular remodeling in PWS [2, 57]. We demonstrated the use of OCT and OCTA to image histopathological and vascular changes in lesion skin and the corresponding normal skin (symmetrical parts of skin lesions on the cheek) from PWS patients. Handheld OCTA offers a user-friendly and fast imaging approach with high repeatability for capturing complex 3D microvascular architecture with high resolution. This makes it a valuable imaging tool for studying the vascular remodeling of PWS lesions. Importantly, OCTA provides detailed vascular maps without skin invasion, preserving the skin's condition for further detection and treatment. Moreover, the majority population of clinical PWS patients is children who have a lower degree of tolerance and compliance under the detection with handheld OCT probe compared to the adults. The SNR-adaptive ID-OCTA algorithm with the inter-frame registration method we used can effectively suppress the motion artifact introduced by the movement of patients, ensuring high imaging quality and reliability in subsequent classification network training and quantitative analysis.

Quantitative OCTA has been instrumental in characterizing diseases in ophthalmology and dermatology. In this work, we introduced six metrics to quantify the angiopathology of PWS lesions. VD, VC, and VX are already mature metrics for quantifying the vascular morphology of PWS. We adopted them after preprocessing operations such as BM3D denoise and Phansalkar binarization. In addition to these three metrics that quantify vascular morphology, the vessels' depth and growth can also influence the distribution and effectiveness of the photosensitizing agent and the light penetration, thereby affecting the therapeutic effects. To quantify the vascular depth information of PWS lesions, we rendered the vascular-depth map in the base of the OCTA image. We acquired two metrics containing the vascular lesions' depth and growth information: MDV and DVG. In generating a vascular-depth map, the upper surface of the cross-sectional OCT image needs to be identified first. Then, another metric, ST, is calculated according to the location of the upper surface. The mean VD of the normal skin in our study is $0.38 \pm 0.06$, which matches the results measured in earlier studies [36-38]. Besides, the phenomenon of no significant difference in vascular morphology metrics between purplish red and thickened types is consistent with previous studies [36]. However, in contrast to other findings where most of the PWS data have higher VD than normal skin data, the proposed classification method suggests that type1 and type 4 PWS data have a decline in VD (Fig. 6a). The inconsistency could be attributable to two possible reasons. Firstly, previous studies had smaller data sizes, all PWS data were calculated together for statistics, and data with declined VD were not analyzed separately. Secondly, type 1 and type 4 PWS may have unique vascular remodeling (e.g., vessel degeneration and atrophy) compared to other types. This hypothesis needs to be further investigated in future works.

Furthermore, the proposed fine-grained classification method of PWS shows an overall increase in VC and a decrease in VX (Fig. 6b-c), which are in accordance with previous studies [36, 37]. The increase of ST (Fig. 6d) also demonstrates the potential of cross-sectional OCT images with identified upper surface as a biomarker of PWS. Only VC and ST present an overall trend of escalation across type 1 to type 5 PWS. This finding indicates that VC and ST can evaluate the severity of PWS, which coincides with former studies that viewed vessel diameter and skin surface condition as the basis of clinical classification of PWS [19, 58]. DVG is a new metric we proposed to test the hypothesis that vessel growth can influence the effect of V-PDT. The decrease of DVG reveals that the vascular network becomes flatter after remodeling and is reflected as a more subtle color change along with the growth of a vessel in the vascular-depth map. Interestingly, only type 4 PWS has an improved DVG compared to the other four types, and this is exactly opposite to the quantification of MDV, where only type 4 PWS has a decreased MDV (Fig. 6e-f). We speculate this surprising result might be due to the irregular growth of blood vessels in type 4 PWS. Blood vessels may grow perpendicularly and curve toward the skin surface in type 4 PWS, leading to a decrease in MDV and an increase in DVG. The particularity of the vascular growth in type 4 PWS may be the key clue to the occasional failure of V-PDT. In the future, we will continue to describe the concrete angiopathological features of each subtype of PWS.

One limitation of the current study is the lack of treatment data about the effect of V-PDT on each subtype of PWS. It is crucial to comprehend the association between the angiopathology of PWS lesions and the effectiveness of V-PDT to develop diagnostic standards, optimize treatment plans, and assess prognosis in clinical practice. Our future study aims to address this gap by comprehensively analyzing metrics derived from OCTA across different types of PWS lesions. Subsequently, we will elucidate the relationship between these metrics and the factors influencing the effectiveness of V-PDT. Drawing inspiration from the applications of radiomics and OCT-omics [59, 60], we propose the development of a diagnosis and treatment system utilizing OCTA-omics. We aim to construct predictive tools by training deep learning models with OCT and OCTA metrics, V-PDT parameters (such as photosensitizer type and dose, light dose), and treatment outcomes (including lesion blanching degree and patient-reported satisfaction). This approach will assist clinicians in devising optimized V-PDT strategies tailored to individual patients with PWS, aligning with the growing trend of personalized medicine. Future research will explore the integration of OCTA-omics with robot-assisted OCTA detection and V-PDT, paving the way for precise therapeutic interventions for a broader spectrum of dermatological conditions.

## V. CONCLUSION

Our study introduces a novel, fine-grained classification method for PWS using OCTA. By developing a quantification framework that evaluates six critical metrics related to vascular morphology and depth information, we demonstrated significant angiopathological differences among PWS subtypes not captured by conventional clinical classifications based on skin appearance. Our results indicate that this new classification method offers a more precise reflection of the heterogeneity in PWS lesions, thereby potentially enhancing the efficacy of V-PDT. This approach is the first attempt to classify PWS lesions directly from OCT and OCTA information, emphasizing the importance of detailed histopathological and vascular characteristics in clinical diagnosis and treatment planning. The significance of our study lies in its potential to revolutionize the management of PWS by providing a non-invasive, reliable quantification to assess angiopathology and a novel perspective to classify PWS. Our findings suggest incorporating OCTA-derived angiopathological insights could lead to more targeted and effective therapeutic strategies, such as OCTA-guided robotic photodynamic and photothermal therapies for dermatology diseases.